\documentstyle[epsf]{mn}

\newif\ifAMStwofonts

\def\kms{\relax \ifmmode {\,\rm km\,s}^{-1}\else \,km\,s$^{-1}$\fi}
\def\ha{\relax \ifmmode {\rm H}\alpha\else H$\alpha$\fi}
\def\hb{\relax \ifmmode {\rm H}\beta\else H$\beta$\fi}
\def\hi{\relax \ifmmode {\rm H\,{\sc i}}\else H\,{\sc i}\fi}
\def\hii{\relax \ifmmode {\rm H\,{\sc ii}}\else H\,{\sc ii}\fi}
\def\h2{\relax \ifmmode {\rm H}_2\else H$_2$\fi}
\def\lha{\relax \ifmmode L_{{\rm H}\alpha}\else $L_{{\rm H}\alpha}$\fi}
\def\shi{\relax \ifmmode \sigma_{{\rm HI}}\else $\sigma_{\rm HI}$\fi}
\def\sh2{\relax \ifmmode \sigma_{{\rm H}_2}\else $\sigma_{{\rm H}_2}$\fi}
\def\degr{\hbox{$^\circ$}}
\def\arcmin{\hbox{$^\prime$}}
\def\arcsec{\hbox{$^{\prime\prime}$}}
\def\deg{\hbox{$^\circ$}}
\def\min{\hbox{$^\prime$}}
\def\sec{\hbox{$^{\prime\prime}$}}
\def\fdg{\hbox{$.\!\!^\circ$}}
\def\fs{\hbox{$.\!\!^{\rm s}$}}
\def\farcm{\hbox{$.\mkern-4mu^\prime$}}
\def\farcs{\hbox{$.\!\!^{\prime\prime}$}}
\def\degd#1.#2{ #1\fdg#2 }                 
\def\mind#1.#2{ #1\farcm#2 }               
\def\secd#1.#2{ #1\farcs#2 }               
\def\hhh{\ifmmode {\rm ^h}              
         \else {${\rm ^h}$}
         \fi}
\def\sss{\ifmmode {\rm ^s}              
         \else {${\rm ^s}$}
         \fi}
\def\hms#1h#2m#3s{                      
                  \relax
                  \ifmmode #1^{\rm h}\,#2^{\rm m}\,#3^{\rm s}
                  \else \hbox{$#1^{\rm h}\,#2^{\rm m}\,#3^{\rm s}$}
                  \fi
                 }
\def\dms#1d#2m#3s{                      
                  \relax
                  #1\degr\,#2\arcmin\,#3\arcsec 
                 }
\def\hmsd#1h#2m#3.#4s{                  
                      \relax
                      \ifmmode #1^{\rm h}\,#2^{\rm m}\,#3\fs#4
                      \else \hbox{$#1^{\rm h}\,#2^{\rm m}\,#3\fs#4$}
                      \fi
                     }
\def\dmsd#1d#2m#3.#4s{                  
                      \relax
                      #1\degr\,#2\arcmin\,#3\farcs#4
                     }
\def\mag{\relax                          
        \ifmmode ^{\rm m}
        \else $^{\rm m}$
        \fi
       }
\def\magd#1.#2{                          
              \relax
              \ifmmode #1^{\rm m}
                       \hskip-0.55em.\hskip0.22em#2
              \else \hbox{#1$^{\rm m}
                    \hskip-0.55em.\hskip0.22em$#2}
              \fi
             }

\ifoldfss
  \ifCUPmtlplainloaded \else
    \NewTextAlphabet{textbfit} {cmbxti10} {}
    \NewTextAlphabet{textbfss} {cmssbx10} {}
    \NewMathAlphabet{mathbfit} {cmbxti10} {} 
    \NewMathAlphabet{mathbfss} {cmssbx10} {} 
  \fi
  \ifAMStwofonts
    \ifCUPmtlplainloaded \else
      \NewSymbolFont{upmath} {eurm10}
      \NewSymbolFont{AMSa} {msam10}
      \NewMathSymbol{\upi}     {0}{upmath}{19}
      \NewMathSymbol{\umu}     {0}{upmath}{16}
      \NewMathSymbol{\upartial}{0}{upmath}{40}
      \NewMathSymbol{\leqslant}{3}{AMSa}{36}
      \NewMathSymbol{\geqslant}{3}{AMSa}{3E}

    \fi
  \fi
\fi 

\ifnfssone
  \newmathalphabet{\mathit}
  \addtoversion{normal}{\mathit}{cmr}{m}{it}
  \addtoversion{bold}{\mathit}{cmr}{bx}{it}
  \newmathalphabet{\mathbfit} 
  \addtoversion{normal}{\mathbfit}{cmr}{bx}{it}
  \addtoversion{bold}{\mathbfit}{cmr}{bx}{it}
  \newmathalphabet{\mathbfss} 
  \addtoversion{normal}{\mathbfss}{cmss}{bx}{n}
  \addtoversion{bold}{\mathbfss}{cmss}{bx}{n}
  \ifAMStwofonts
    \ifCUPmtlplainloaded \else
      \UseAMStwoboldmath
      \makeatletter
      \new@mathgroup\upmath@group
      \define@mathgroup\mv@normal\upmath@group{eur}{m}{n}
      \define@mathgroup\mv@bold\upmath@group{eur}{b}{n}
      \edef\UPM{\hexnumber\upmath@group}
      \new@mathgroup\amsa@group
      \define@mathgroup\mv@normal\amsa@group{msa}{m}{n}
      \define@mathgroup\mv@bold\amsa@group{msa}{m}{n}
      \edef\AMSa{\hexnumber\amsa@group}
      \makeatother
      \mathchardef\upi="0\UPM19
      \mathchardef\umu="0\UPM16
      \mathchardef\upartial="0\UPM40
      \mathchardef\leqslant="3\AMSa36
      \mathchardef\geqslant="3\AMSa3E
    \fi
  \fi
\fi 

\ifnfsstwo
  \DeclareMathAlphabet{\mathbfit}{OT1}{cmr}{bx}{it}
  \SetMathAlphabet\mathbfit{bold}{OT1}{cmr}{bx}{it}
  \DeclareMathAlphabet{\mathbfss}{OT1}{cmss}{bx}{n}
  \SetMathAlphabet\mathbfss{bold}{OT1}{cmss}{bx}{n}
  \ifAMStwofonts
    \ifCUPmtlplainloaded \else
      \DeclareSymbolFont{UPM}{U}{eur}{m}{n}
      \SetSymbolFont{UPM}{bold}{U}{eur}{b}{n}
      \DeclareSymbolFont{AMSa}{U}{msa}{m}{n}
      \DeclareMathSymbol{\upi}{0}{UPM}{"19}
      \DeclareMathSymbol{\umu}{0}{UPM}{"16}
      \DeclareMathSymbol{\upartial}{0}{UPM}{"40}
      \DeclareMathSymbol{\leqslant}{3}{AMSa}{"36}
      \DeclareMathSymbol{\geqslant}{3}{AMSa}{"3E}
    \fi
  \fi
\fi 

\ifCUPmtlplainloaded \else
  \ifAMStwofonts \else 
    \def\upi{\pi}
    \def\umu{\mu}
    \def\upartial{\partial}
  \fi
\fi

\title[H{\sc i} in NGC 3631]{Atomic hydrogen in the spiral galaxy NGC 3631}
\author[J. H. Knapen]
       {J. H. Knapen$^{1,2}$\\
$^1$Department of Physical Sciences, University of Hertfordshire, 
College Lane, Hatfield, Herts AL10 9AB, UK (present address)\\ E-mail
knapen@star.herts.ac.uk\\
$^2$D\'epartement de Physique, Universit\'e de Montr\'eal, C.P.
6128, Succursale Centre-Ville, Montr\'eal (Qu\'ebec), H3C 3J7 Canada}

\date{Accepted 12 November 1996.
      Received;
      in original form}

\pagerange{\pageref{firstpage}--\pageref{lastpage}}
\pubyear{1996}

\begin{document}

\maketitle

\label{firstpage}

\begin{abstract}

New high resolution, high sensitivity WSRT \hi\ synthesis observations
of the spiral galaxy NGC~3631 are presented. In the total atomic
hydrogen map, the spiral arms are well distinguished from the interarm
regions, while the sensitivity allows detection of \hi\ in all but a
few isolated regions of the areas between the spiral arms. Most of the
atomic hydrogen is located within the optical disc, but the \hi\
extends to some $1.5\times R_{opt}$. The \hi\ follows the spiral arms,
and streaming motions of up to $\sim15$\,\kms\ (projected) can be
identified from the velocity field.  Assuming a constant inclination
angle of 17\deg, a rotation curve is derived which is slightly falling
in the outer parts of the disc.  Analysis of a residual velocity
field, obtained after subtraction of an axisymmetric model based on
the rotation curve, confirms the existence of streaming motions near
the spiral arms in an otherwise undisturbed disc.

\end{abstract}


\section{Introduction}

In the study of star formation (SF) processes in discs of spiral
galaxies, it seems natural to distinguish explicitly between arm and
interarm environments. This is however not usually done, partly
because of the problems such an approach entails regarding the spatial
resolution of especially radio and molecular observations. But also in
the optical regime, where it is easy to isolate spiral arms on images
of nearby galaxies, photometric properties are usually studied by
means of azimuthally averaged profiles. Knapen \& Beckman (1996) and
Beckman et al. (1996) used optical images to derive radial profiles
and scale lengths for the spiral arm and interarm regions separately
for 3 galaxies: M51, M100=NGC~4321 and NGC~3631, and found significant
differences between arm and interarm profiles, with larger scale
lengths in the arms in all three galaxies considered. For M100, Knapen
\& Beckman (1996) also included \ha, \hi, radio continuum and CO
observations, and concluded that the shape of the radial (whole disc,
and arm/interarm) profiles is determined by SF more than by any other
factor. Knapen \& Beckman noted in particular that the \hi\ is
enhanced in the region of the star-forming spiral arms, which they
interpret as a result, through photodissociation of part of the
molecular gas, rather than as a cause of the SF.

In another line of work, massive star formation efficiencies (MSFE)
along the spiral arms are compared directly with values in the adjacent
interarm regions (Cepa \& Beckman 1990; Knapen et al. 1992, 1996). Such
a comparison can only be made using CO, \hi, and \ha\ data of sufficient
spatial resolution to isolate the spiral arms. For M51 and M100, two
galaxies where observations of these three tracers could be used,
enhancement of the efficiencies along the spiral arms could be directly
interpreted as evidence for triggering of the SF in the arms (Knapen et
al. 1992, 1996).

In order to study the r\^{o}le of atomic hydrogen and the interplay
between gas and stars at the scale of spiral arms, \hi\ observations at
resolutions equivalent to at most the width of a spiral arm are
needed. In a typical galaxy at a distance of $\sim15$ Mpc, such as M100,
M51, or in fact NGC~3631, this implies the need for a spatial resolution
of $\sim15\sec$. Since the \hi\ emission from the interarm is generally
a factor 3--5 lower than that from the arms, the need for observing the
interarm \hi\ also implies that observations of high sensitivity are
needed. The observations described in the present paper were designed
precisely to meet these goals: to reliably measure interarm \hi\
emission over the disc of NGC~3631.

Candidate spiral galaxies for a study of the SF processes in their
arms and disc should be objects with well-defined spiral arms, usually
of late morphological type. They should also be relatively face-on
objects, with say $i<30\deg$, since projection effects and reduced
effective spatial resolution make it much more difficult to
distinguish spiral arms in more inclined galaxies. This is a class of
spiral galaxies not traditionally studied in great detail through \hi\
synthesis observations, since the velocity information that can be
deduced for such low-inclination galaxies is necessarily limited. In
our \hi\ study of M100 ($i\sim27\deg$) we had to adopt and fix an
inclination angle when calculating a rotation curve (Knapen et
al. 1993), and I will have to repeat that procedure in the present
study.

We have chosen the grand-design galaxy NGC~3631 for our \hi\ study. We
follow the approach followed before for M100, where the \hi\ data
(Knapen et al. 1993) were used for detailed studies of efficiencies,
SF processes, and for determining the location of the density wave
resonances in the disc (Knapen et al. 1996; Knapen \& Beckman 1996;
Sempere et al. 1995, respectively). In forthcoming papers, we will
describe similar work for NGC~3631, using the \hi\ data as presented
in this paper.

NGC~3631 was first observed at radio wavelengths by Roberts
(1968). Other single-dish \hi\ observations include those by Fisher \&
Tully (1981), Tifft \& Cocke (1988) and Staveley-Smith \& Davies
(1988). No synthesis observations have been described so far. NGC~3631
is a late-type (Sc), face-on spiral galaxy, looking conspicuously
``normal'' at optical wavelengths. It is non-barred, has no obvious
companions, and shows no other signs of important dynamical
perturbations. The distribution of \hii\ regions in the disc was
described by Boeshaar \& Hodge (1977), who also studied the spiral arm
shape. Recently, we obtained a new high-quality \ha\ image of the
galaxy, from which we catalogued more than 1300 individual \hii\ regions
(Rozas, Beckman \& Knapen 1996).

After discussing the details of the radio and optical observations and
data reduction in Sect.~2, the distribution of \hi\ at different
resolutions is described in Sect.~3.  Sect.~4 is devoted to the
kinematics, and includes a derivation of the rotation curve from the
velocity field. The main results of the paper are briefly summarized in
Sect.~5.

\section{Observations and reduction}

\subsection{Atomic hydrogen}

A field centred on NGC~3631 was observed in the atomic hydrogen 21~cm
line with the Westerbork Synthesis Radio Telescope (WSRT) during the
months of April and May, 1991. The total observing time of 44 hours
was made up out of two periods of 12\,h, one of 9\,h, one of 8\,h and
one of 3\,h. The total bandwidth of the observations was 2.5~MHz,
centred at a heliocentric velocity of 1160 \kms, and divided into 128
evenly spaced channels of 4.13 \kms\ each. The observational
parameters are fully detailed in Table~1. The data were reduced using
the Newstar programs. Since there were many continuum sources within
the field, 20 of them were subtracted from the UV data. The data were
then Fourier transformed to a 512$\times$512 grid of \secd 4.0
$\times$ \secd 6.0 ($\alpha\times\delta$) pixels. The resulting
128-channel datacube was Hanning-smoothed along the velocity
coordinate, resulting in a dataset with a velocity resolution of 8.25
\kms.

\begin{figure*}
  \epsfxsize=18cm \epsfbox{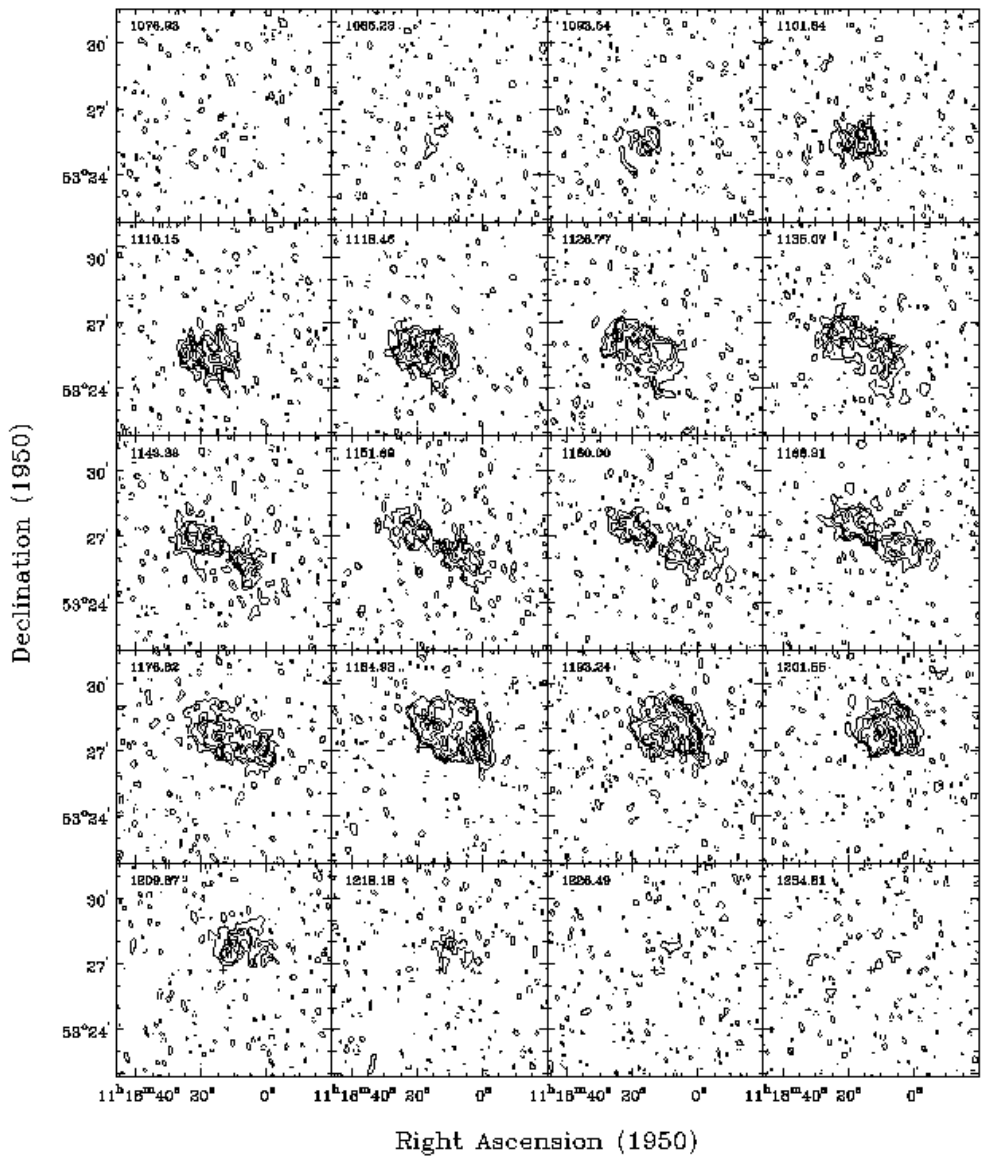}
  \caption{({\it a}) Channel maps of the continuum-subtracted \hi\ line
  emission for NGC~3631, at 21\sec$\times$14\sec\ resolution. Contour
  levels are $-3.8$, $-1.9$ (dashed), 1.9 ($=2\sigma$), 3.8, 5.7, 8.6,
  and 11.4 mJy\,beam$^{-1}$. Only every second channel is shown. The
  centre of NGC~3631 is indicated with a cross. The heliocentric
  velocity of each map is shown in the upper left-hand corner. The beam
  size is indicated as a hatched ellipse in the bottom-left corner of
  the last panel.}
\end{figure*}

\setcounter{figure}{0}

\begin{figure*}
  \epsfxsize=18cm \epsfbox{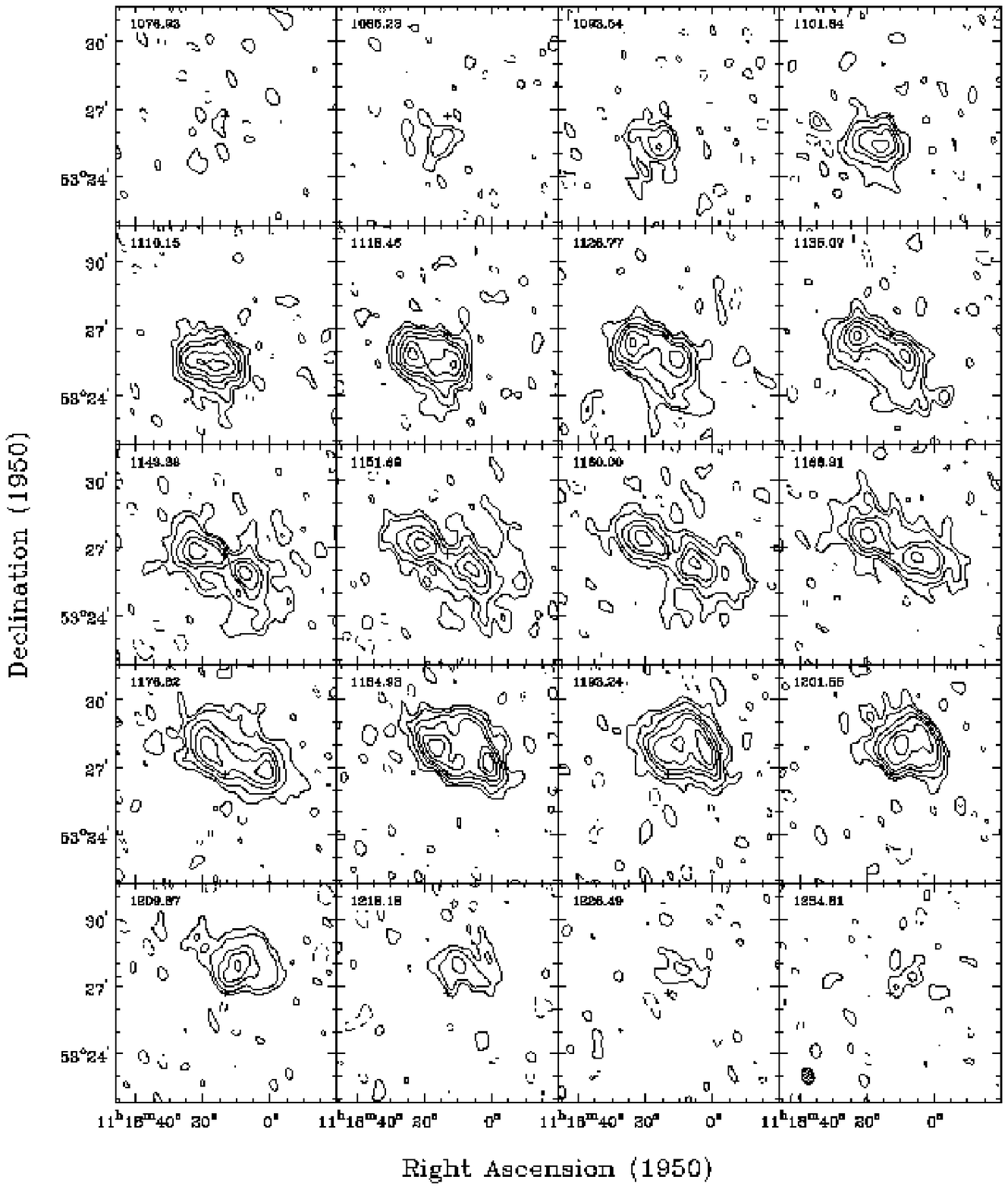}
  \caption{({\it b}) Channel maps at 45\sec$\times$30\sec\ resolution.
  Contour levels are $-5.5$, $-2.8$ (dashed), 2.8 ($=2\sigma$), 5.5,
  11.1, 16.6, 24.9, 33.2, and 49.9 mJy\,beam$^{-1}$. Only every second
  channel is shown. The centre of NGC~3631 is indicated with a
  cross. The heliocentric velocity of each map is shown in the upper
  left-hand corner. The beam size is indicated as a hatched ellipse in
  the bottom-left corner of the last panel.}
\end{figure*}

\begin{table}
 \centering
 \begin{minipage}{140mm}
  \caption{Observing parameters}
\begin{tabular}{l l}
\hline\hline
Galaxy & 			NGC 3631\\
Telescope &			WSRT\\
Date of observation &		April/May 1991\\
Duration of observation &	44 h\\
Number of interferometers &	40\\
Baselines (min-max-incr) &	36--2700--72\\
Synthesized beam (FWHM) & 	\secd 15.2 $\times$ \secd 11.2\\
Pos. angle synth. beam &	\degd 12.7\\ 
FWHP primary beam &		37\min\\
Velocity central channel &	1160.0 \kms\ (heliocentric)\\
Bandwidth &			2.5 MHz\\
Number of channels &		128\\
Channel separation &		4.13 \kms\\
Velocity resolution &		8.25 \kms\\
Field centre (1950) &		($\!$\hmsd 11h18m13.2s; \dms 53d26m43s)\\
\hline\hline
\end{tabular}
\end{minipage}
\end{table}

After convolving the datacube to a resolution of some 60\sec, 28
channels free of line emission were identified on the low-velocity
side, and 23 channels in the high-velocity side. A number of channels
with higher noise was discarded on either side.  Since we are
interested in structure at several scales and sensitivities, the
data-cube was convolved from the original resolution of $\secd
15.2\times\secd 11.2$ to resolutions of 21\sec$\times$ 14\sec,
30\sec$\times$ 20\sec, 45\sec$\times$ 30\sec\ and 60\sec$\times$
40\sec.  The continuum was determined for each of these data sets by
fitting a linear relation to the line-free channels, and subtracted
from the line emission channels.  The resulting five data sets were
cleaned.  Clean components were subtracted until the noise level, and
subsequently the maps were restored by convolving the components with
the appropriate Gaussian beam and adding the residuals.  Five data
cubes were thus produced, consisting of 58 line channels each, at
resolutions ranging from $\secd 15.2\times\secd 11.2$ (hereafter full
resolution) to 60\sec$\times$ 40\sec. The cleaned channel maps at
21\sec$\times$ 14\sec\ and 45\sec$\times$ 30\sec\ resolution are shown
in Figures 1a and 1b, respectively. Note that only every second
channel map is shown.  Noise properties of these datacubes and
conversion factors $T_b({\rm K})/S({\rm mJy})$ (equivalent $T_b$ of 1
mJy\,beam$^{-1}$) are listed in Table~2.

The data cubes at various resolutions were used to calculate total
\hi\ (zero-th moment), velocity (first moment) and velocity dispersion
(second moment) maps. A careful inspection of the data cubes showed
that such an analysis is in fact valid over almost the complete disc,
with the possible exception of the central region (of size
approximately one beam), where the \hi\ profiles are not
Gaussian-shaped and/or not single-peaked.  I now briefly describe the
procedure followed for the full resolution data set.  The first step
was to produce a conditionally transferred data cube, in which values
were only retained at positions where the intensity in the smoothed
data cube at 30\sec$\times$ 20\sec\ was larger than 2.5 times the rms
noise of the smoothed maps. Pixel values at all other positions were
set to undefined. Then, noise peaks outside the area where \hi\
emission is expected were removed by setting pixel values at those
positions to undefined. This was done interactively by inspecting the
(high resolution) channel maps one by one, continually comparing with
the same channel and referring to adjacent channels in the smoothed
cube (at 30\sec$\times$ 20\sec). Finally, the resulting data set was
used as input for the GIPSY program {\sc moments} to calculate the
total intensity, velocity and velocity dispersion maps. Only emission
occurring at the same position in at least three adjacent channels was
considered true signal, and used for the calculation of the moment
maps.

The procedure for making the moment maps of the lower-resolution data
sets is completely analogous. For the 21\sec$\times$ 14\sec\ resolution
data set, for instance, the smoothed cube at 45\sec$\times$ 30\sec\ was
used for reference. A data set of 90\sec$\times$ 60\sec\ was used for
reference for both the 45\sec$\times$ 30\sec\ and 60\sec$\times$ 40\sec\
cubes, but in those cases additional restrictions were employed in the
moment calculation, where only values of $>2.5\sigma$ (and
$<-2.5\sigma$) were used.

\begin{table}
 \centering
 \begin{minipage}{140mm}
  \caption{Map properties}
\begin{tabular}{ccc}
\hline\hline
Synthesized Beam  & rms Noise in & Conversion factor \\
(FWHM) &	   channel maps & $T_b({\rm K})/S(${\rm mJy}) \\
 & 			(mJy\,beam$^{-1}$) & \\
\hline
$\secd 15.2\times\secd 11.2$		& 0.84 & 3.59\\
$\secd 21.0\times\secd 13.9$		& 0.95 & 2.09\\
$\secd 30.1\times\secd 19.9$		& 1.11 & 1.02\\
$\secd 45.1\times\secd 29.9$		& 1.39 & 0.45\\
$\secd 60.0\times\secd 40.0$		& 1.60 & 0.25\\
\hline\hline
\end{tabular}
\end{minipage}
\end{table}

\subsection{Optical imaging}

Images in the $B, V, R$ and $I$ broad-bands were obtained in service
time with the 1m Jacobus Kapteyn Telescope (JKT) on La Palma, on
Dec. 20, 1993. An EEV CCD chip was used of 1242$\times$1152 pixels of
\secd 0.{31} projected size, giving a field of view of around 6
arcmin. The raw images were bias-subtracted and flat-fielded using dawn
sky exposures, and photometrically calibrated using standard stars
observed during the night. The resolution (seeing) in the reduced images
is around \secd 1.5. The $B$ image as used in the present paper has a
pixel scale of \secd 0.{62}. The position of foreground stars in the
images was used to place the images on a correct RA-dec grid. Since the
present images were too small to contain a sufficient number of bright
stars to warrant a good astrometrical solution, they were compared with
the \ha\ image of Rozas et al. (1996), which has a larger field of view,
and for which satisfactory astrometry could be performed using star
positions from the HST GSC. The resulting error in the astrometry of the
optical images as used here is less than $\secd 0.2\!$.  We show the
$B$-band image of NGC~3631 in a gray-scale representation in Fig.~3.

\section{Distribution of \hi\ and continuum emission}

\subsection{Global \hi\ properties}

\begin{figure}
  \epsfxsize=8.5cm \epsfbox{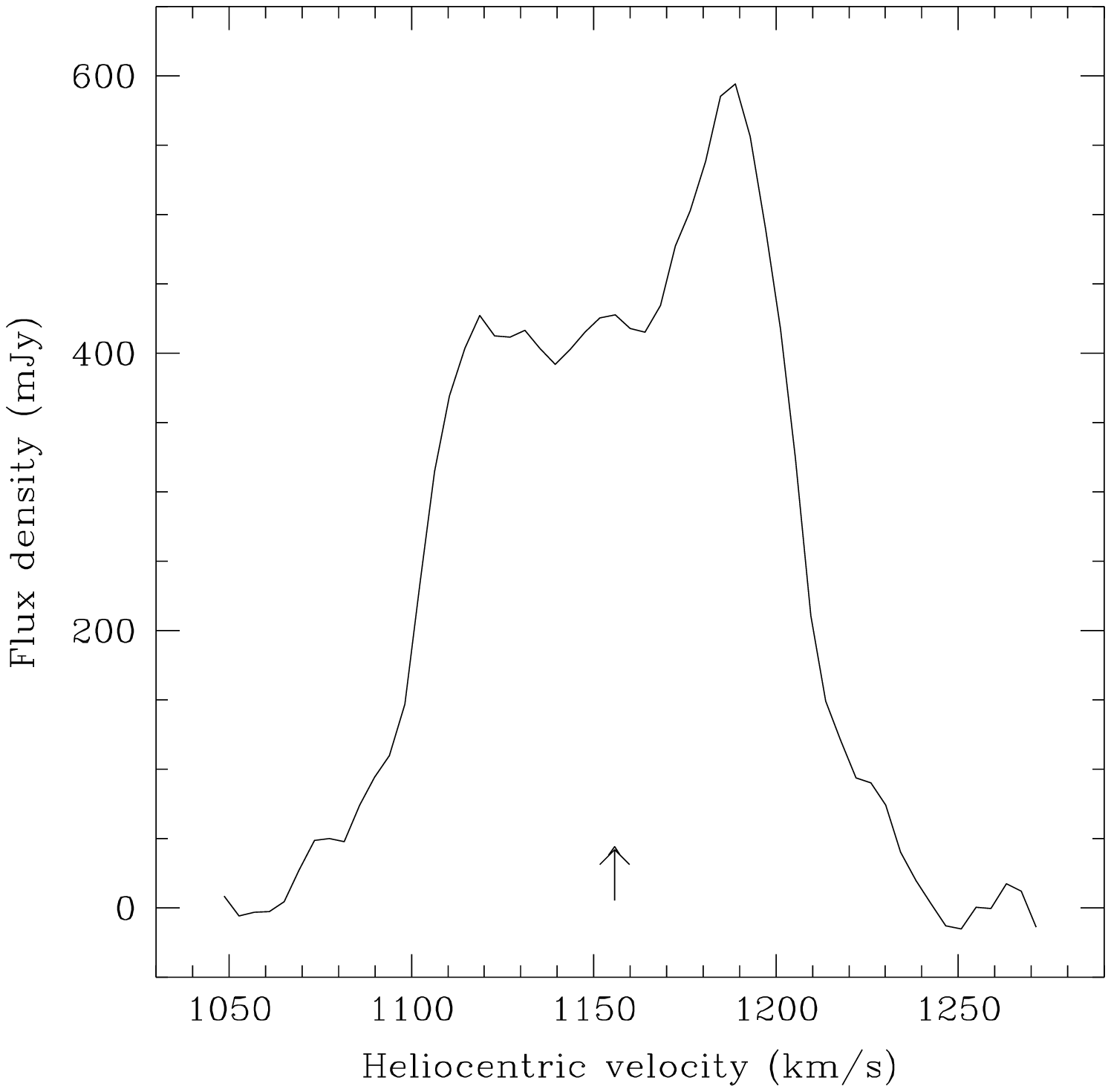}
  \caption{Global \hi\ profile of NGC~3631, as derived from the
  30\sec$\times$ 20\sec resolution data set. Flux densities are
  corrected for primary beam attenuation. The arrow indicates the
  systemic velocity ($v_{sys}=1155.7$ \kms).}
\end{figure}

The global \hi\ line profile is shown in Fig.~2. It was produced by
adding all the flux in each separate channel map of the 30\sec$\times$
20\sec resolution data set, within a square region encompassing the
extent of the \hi\ emission. The systemic velocity (see Sect. 4) of
1155.7\kms\ is indicated by an arrow in the figure. The \hi\ flux
integral is $\int S\,dv=51.6\pm2$\,Jy\,\kms, in good agreement with
the single-dish values given by Tifft \& Cocke (1988) of
54.50~Jy\,\kms\ (no error indicated), and Staveley-Smith \& Davies
(1988) of $50.7\pm4.9$\,Jy\,\kms. Using a distance to NGC~3631 of
15.4~Mpc (derived from $v_{sys}$ assuming $H_0=75$\kms\,Mpc$^{-1}$),
the total atomic hydrogen mass can then be evaluated as $M$(\hi)$=2.9
(\pm0.1) \times10^9M_\odot$. Note, however, that this value does
depend on the true distance to the galaxy, and also on the assumption
that all the atomic hydrogen is optically thin.  The new value
compares favourably to the value reported by Fisher \& Tully (1981) of
$M$(\hi)$=2.9\times10^9M_\odot$ (corrected to $D=15.4$\,Mpc), but not
to the older value of $M$(\hi)$=4.5\times10^9M_\odot$ (also at
$D=15.4$\,Mpc) from Roberts (1968). In general, one can state that the
integrated \hi\ flux as derived from the WSRT synthesis observations
agrees very well with previous determinations made using single-dish
telescopes.

\subsection{Total \hi\ distribution}

\begin{figure*}
  \epsfxsize=18cm \epsfbox{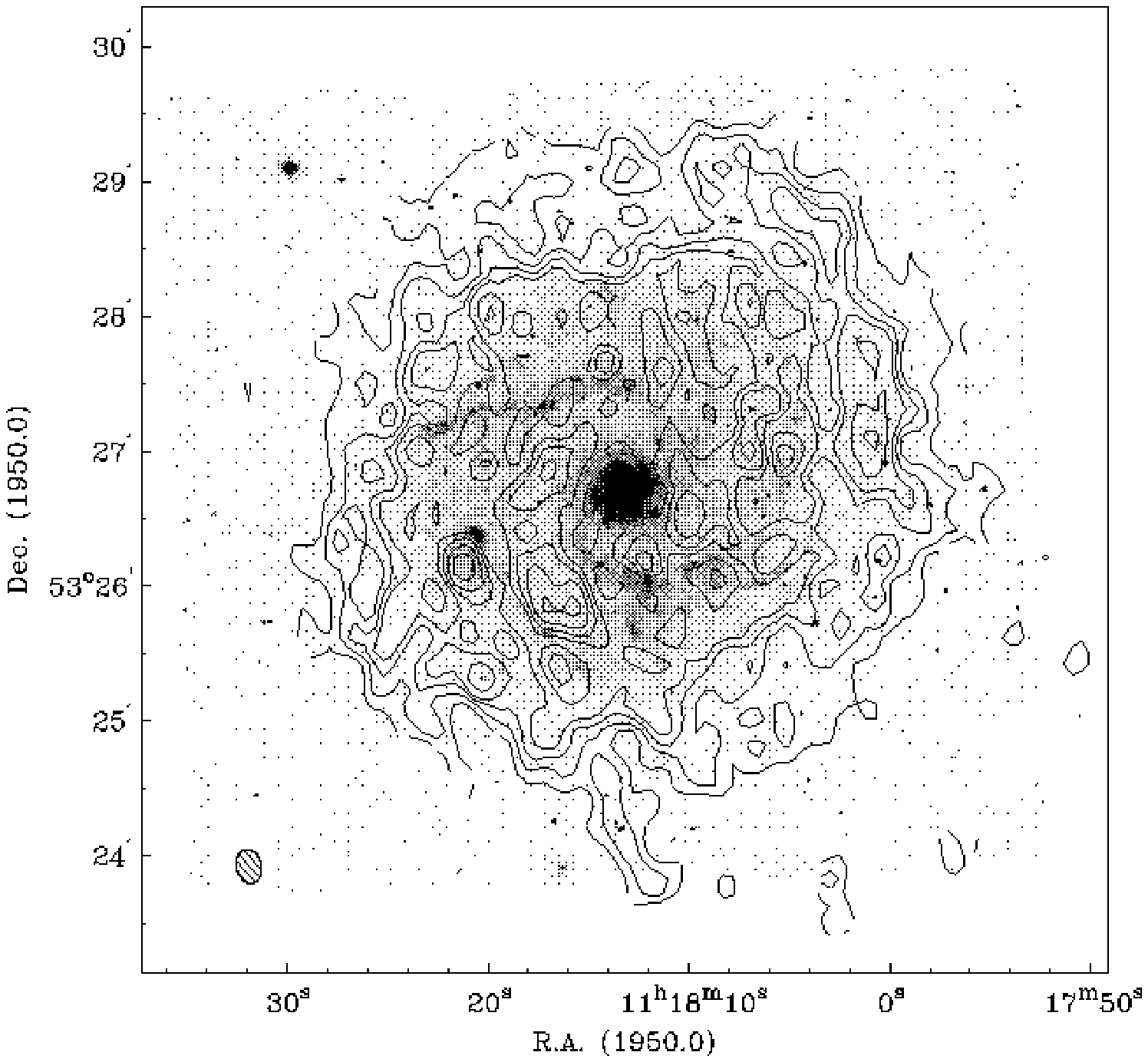}
  \caption{$\secd 15.2\times\secd 11.2$ total \hi\ column density map
  (contours) overlaid on a grayscale representation of a $B$-band CCD
  image of NGC~3631. Contour levels are 2.0, 4.1, 6.1, 8,2, 12.2 and
  16.3$\times10^{20}$\,atoms\,cm$^{-2}$. Beam size is indicated. Note
  that the local minima in the \hi\ distribution can be recognised by
  comparison with the grey-scale representation of the same \hi\ map
  in Fig.~4a.}
\end{figure*}

Figure~3 is an overlay of the $\secd 15.2\times\secd 11.2$ total \hi\
(zero-th moment), or column density of hydrogen, map of NGC~3631 on the
$B$-band CCD image.  The \hi\ generally traces the spiral arms, although
the correspondence between optical and \hi\ features is not unique.  The
\hi\ extends further out than the optical disc, especially in the SW
region of the \hi\ extension (see below and Fig.~4b), but it does not do
so in the dramatic fashion as seen in e.g.  M101 (Allen et al.  1974;
van der Hulst \& Sancisi 1988) or NGC~628 (Kamphuis \& Briggs 1992), two
galaxies of morphological type comparable to NGC~3631. 

\begin{figure*}
  \epsfysize=10.5cm \epsfbox{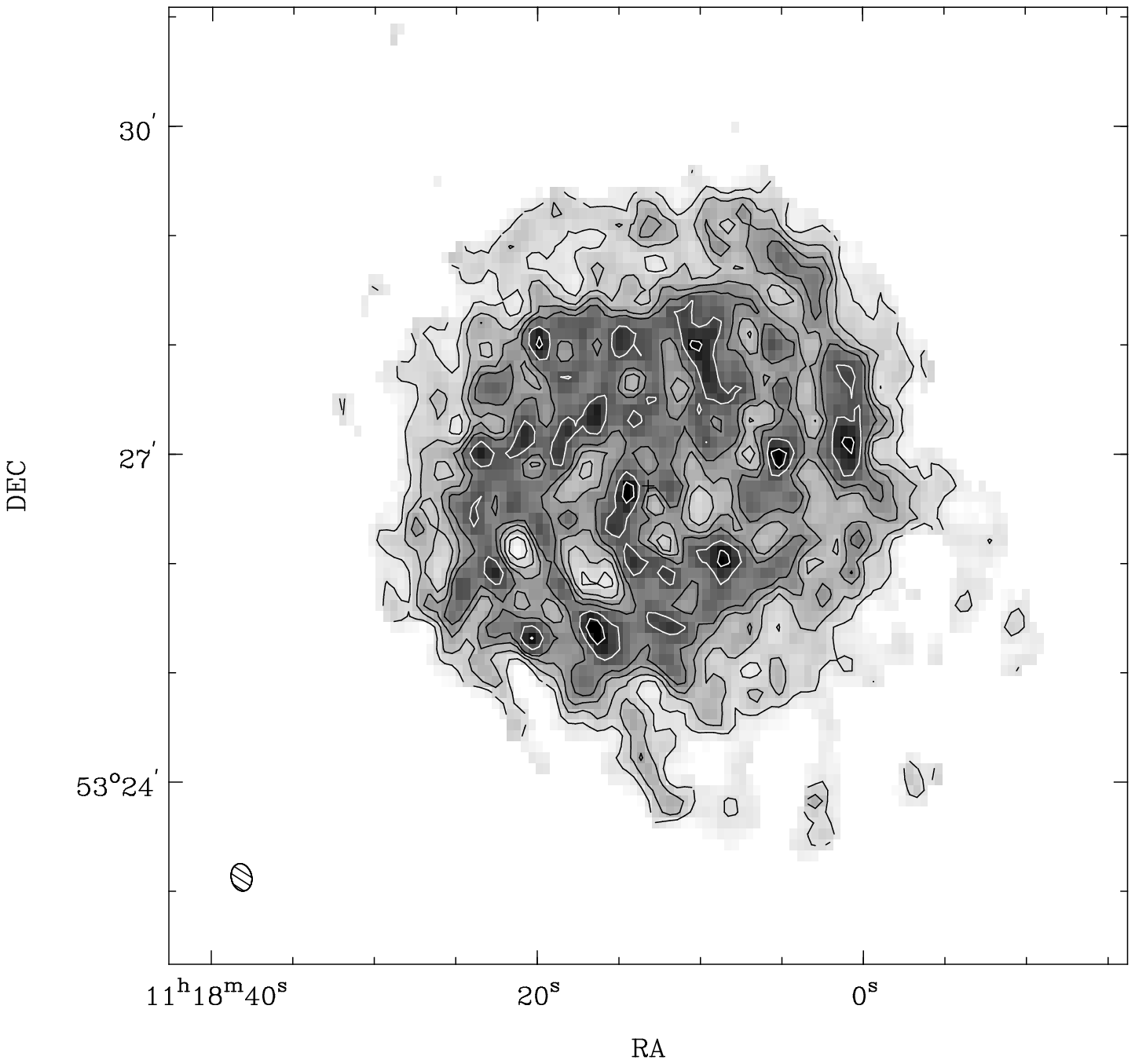}
  \epsfysize=10.5cm \epsfbox{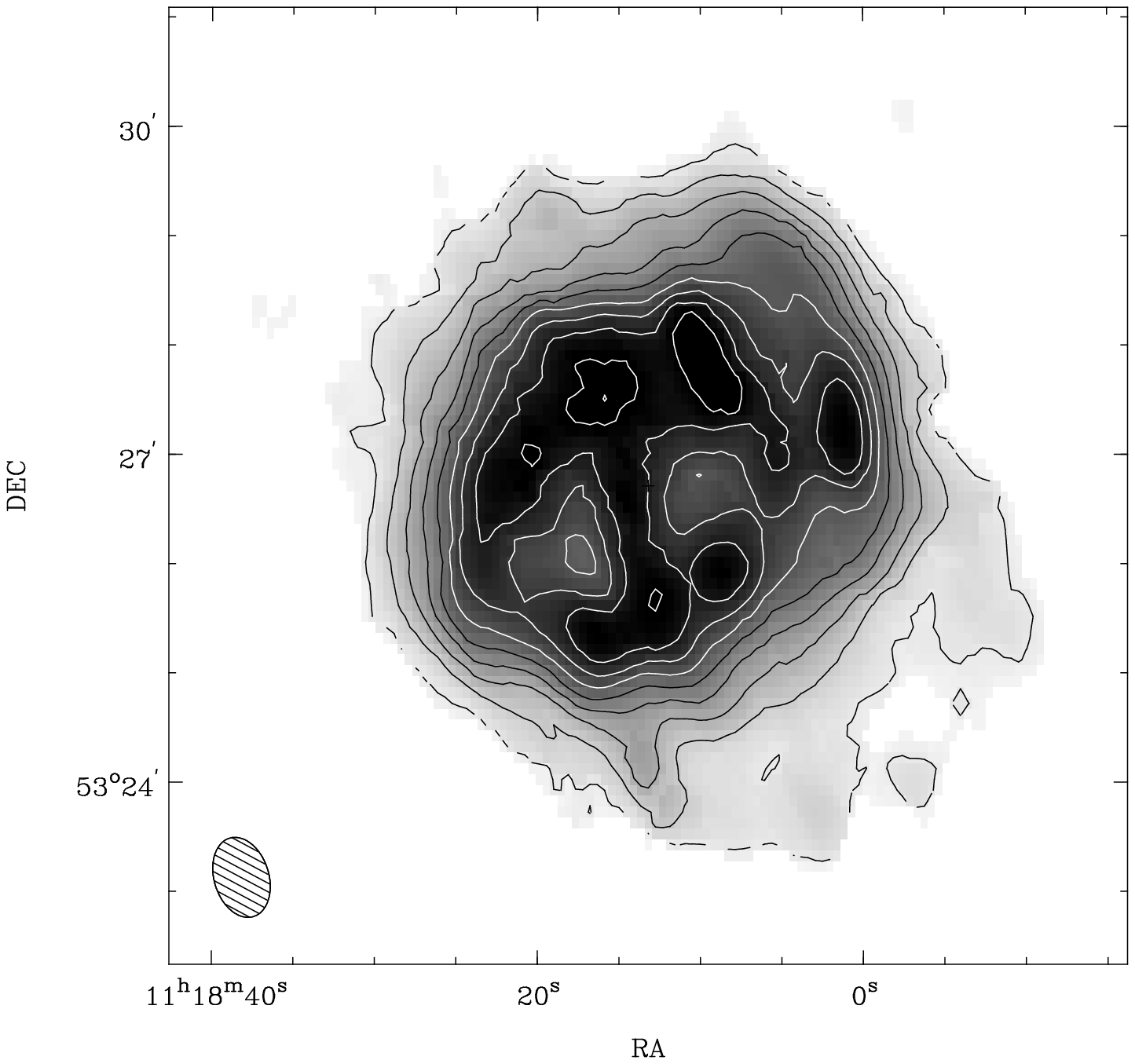}
  \caption{{\it a} (upper) Total \hi\ (column density) map of NGC~3631
  at $\secd 15.2\times\secd 11.2$ resolution. Contour levels as in
  Fig.~3. The centre is indicated with a cross, and the beam size is
  indicated in the lower left corner. Outermost contour is interrupted
  by undefined points. {\it b} (lower) As Fig.~4a, now at
  45\sec$\times$ 30\sec resolution. Contour levels are from 0.85 to
  10.37 in steps of 1.19$\times10^{20}$\,atoms\,cm$^{-2}$. Beam size
  is indicated.}
\end{figure*}

Figures 4a and 4b show the \hi\ column density distribution (zero-th
moment maps) in NGC~3631 at resolutions of $\secd 15.2\times\secd 11.2$
and 45\sec$\times$ 30\sec, respectively.  The higher resolution map
(Fig.~4a) shows that the \hi\ is mostly concentrated in the spiral arms,
but that \hi\ is detected all over the disc, including the interarm
zones and the central region of the galaxy.  At 45\sec$\times$ 30\sec
resolution (Fig.~4b) the general arm pattern can still be recognized but
is mostly lost in the lower resolution.  An interesting feature is the
extension on the SW side of the disc, at low levels.  This is
reminiscent of the distribution of \hi\ in M100, but is not accompanied
by a kinematic signature, as is the case in M100.  Similar \hi\
asymmetries were first recognised by Baldwin, Lynden-Bell and Sancisi
(1980), and seem to be a frequent phenomenon. 

\subsection{Radial \hi\ profile}

\begin{figure}
  \epsfxsize=8.5cm \epsfbox{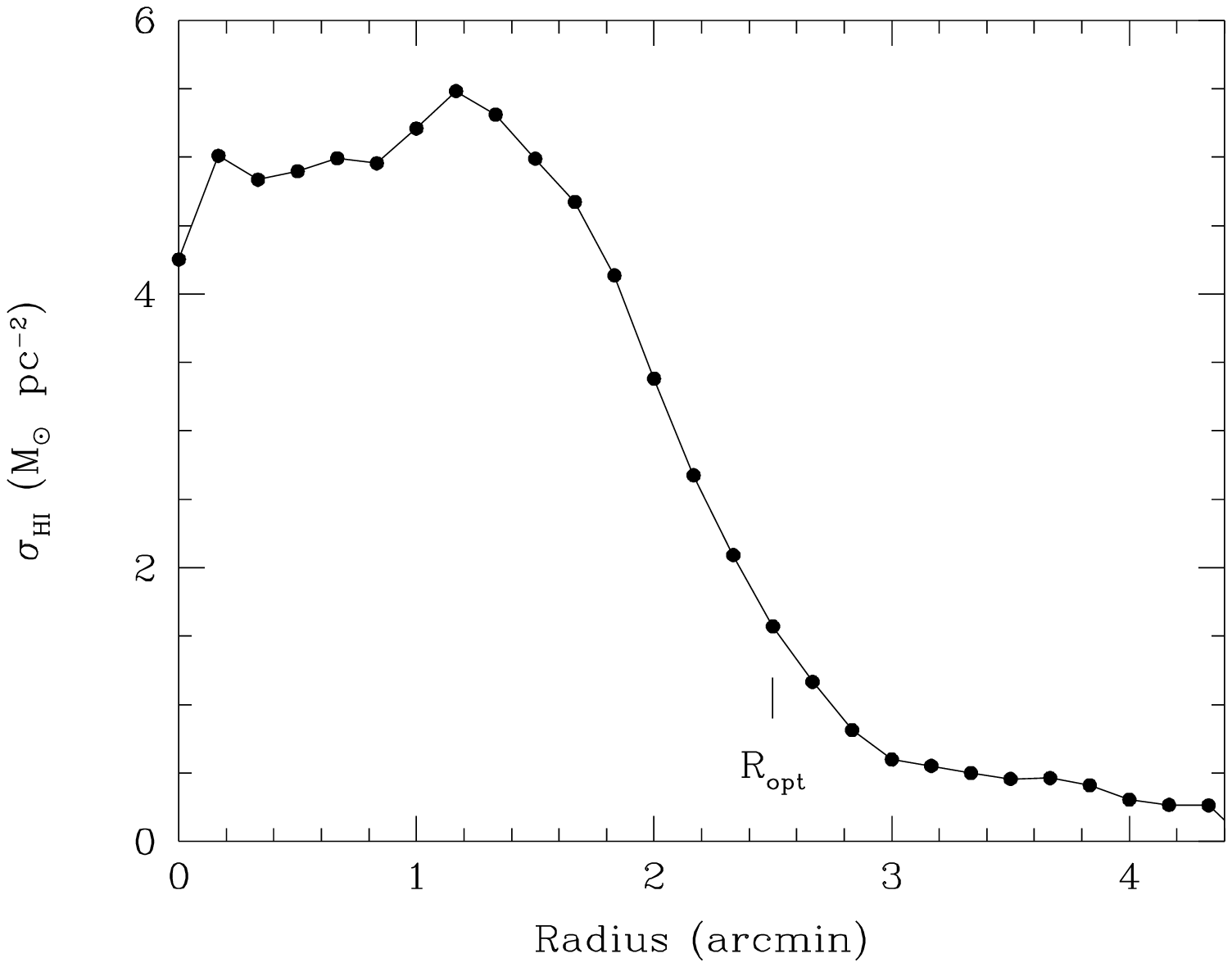}
  \caption{Radial \hi\ surface density profile of NGC~3631, as derived
  from the 30\sec$\times$ 20\sec total \hi\ map by averaging in
  elliptical rings of fixed inclination and position angles. The
  extent of the optical disc ($R_{opt}$) is indicated. Note that the
  \hi\ outside $R\sim3\min$ resides in the SW extension, rather than
  being distributed in the disc (see text).}
\end{figure}

A radial \hi\ surface density profile was derived from the
30\sec$\times$ 20\sec\ total \hi\ map by averaging in elliptical rings
with fixed inclination and position angles ($i=17\deg$ and PA$=150\deg$
respectively, see Sect.~4), and is shown in Fig.~5.  The extent of the
optical disc ($R_{opt}=(0.5\times D_{25})$; from de Vaucouleurs et al. 
1991) is indicated in the Figure.  The profile shows clearly some
features that were already obvious from the overlay of the total \hi\
map on the $B$-band optical image (Fig.~3), namely that the \hi\ disc is
not a whole lot more extended than the optical disc, and that most of
the \hi\ sits well inside the optical disc.  The radial \hi\ profile
shows a central depression, not uncommon at all for both barred and
non-barred galaxies (see e.g.  Broeils \& van Woerden 1994).  The
profile peaks at a radius of just over 1 arcmin, and falls off rapidly
after that radius, coinciding with the end of the region of the star
forming spiral arms as seen in the optical image.  This situation is
reminiscent of that in M100, where also the \hi\ is enhanced in the
region of the SF spiral arms.  Knapen \& Beckman (1996) interpret this
enhancement in M100 as a result of the SF activity, leading to
photodissociation of part of the molecular gas and to the production of
\hi, rather than as the origin of the SF.  Careful comparison with
especially molecular gas observations is needed to confirm that a
scenario of \hi\ production as a result of SF activity is in fact the
preferred one also for NGC~3631. 

The radial \hi\ profile extends out to around $R=4\min$, but
inspection of especially Fig.~4b shows that the \hi\ only extends to
this radius in the SW, elsewhere the disc can not be defined for radii
larger than $R\sim3\min$.  This implies that the surface density
beyond 3 arcmin in Fig.~5 is no longer a ring average, as in the inner
parts, but is due mostly or even exclusively to the \hi\ in the SW
extension.  Comparison of the whole radial \hi\ profile of NGC~3631
with those for other Sc galaxies (Broeils \& Rhee 1996) shows that the
\hi\ disc in NGC~3631 is rather small.  For NGC~3631 we find a value
for the ratio of \hi\ to $B$ diameter $D_{HI}/D_{25}$ of 1.1 (using
$\sigma_{hi}=1\,$M$_\odot$\,pc$^{-2}$ for the definition of $D_{HI}$).
Broeils \& Rhee (1996) find that these ratios for the around 15 Sc
galaxies in their sample are between 1.1 and 2.5, placing NGC~3631 at
the lower extreme.  The peak surface density in NGC~3631, of some
7\,M$_\odot$\,pc$^{-2}$, and the average surface density of around
4.5\,M$_\odot$\,pc$^{-2}$, are quite normal when compared to both the
Sc galaxies, and also to the galaxies of other morphological types,
from Broeils \& Rhee (1996).

\subsection{21~cm Continuum}

\begin{figure}
  \epsfxsize=8.5cm \epsfbox{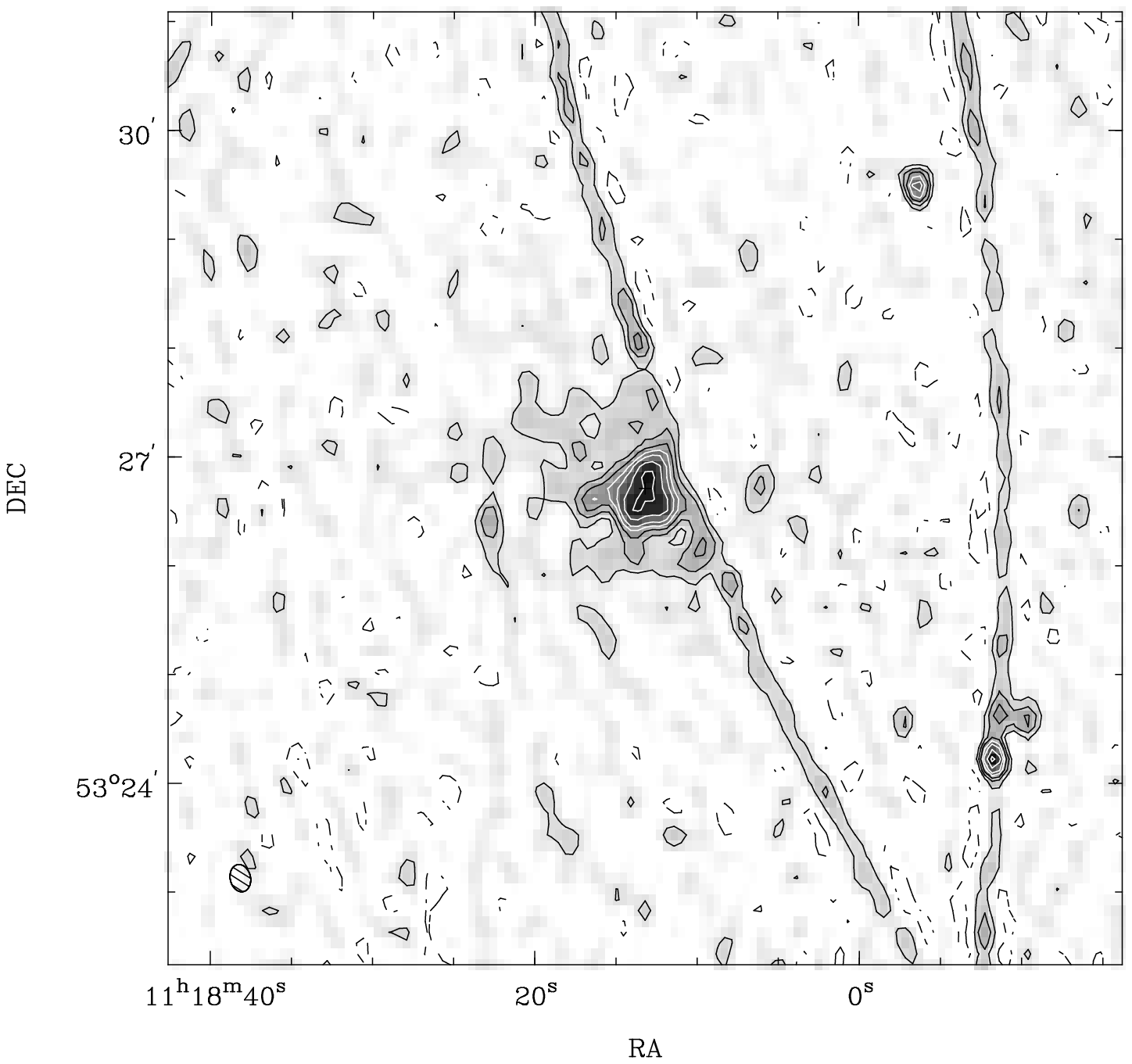}
  \caption{$\secd 15.2\times\secd 11.2$ 21~cm continuum map (contours
  and grayscales). Contour levels are $-1.0$, $-0.5$ (dashed), 0.5
  ($=2\sigma$), 1.0, 1.5, 2.0, 2.5, 3.3 and 4.0 mJy\,beam$^{-1}$. Beam
  size is indicated.}
\end{figure}

The 21~cm continuum map, at $\secd 15.2\times\secd 11.2$ resolution,
is shown in Fig.~6. This is the image that was subtracted from the
data cube of the same resolution. As described in Sect.~2, a number of
continuum sources were deleted from the data set at the stage
preceding the making of the channel maps, but still a number of
grating rings are present in the final continuum map. This makes
interpretation of the features in the map, and especially the
determination of fluxes, very difficult. What can be deduced from the
map is that the 21~cm continuum emission from NGC~3631 is centrally
peaked, and follows the spiral arm shape. No obvious radio continuum
point sources are seen in the disc, thus no recent supernova events
can be identified.

\section{Kinematics}

\begin{figure*}
  \epsfxsize=18cm \epsfbox{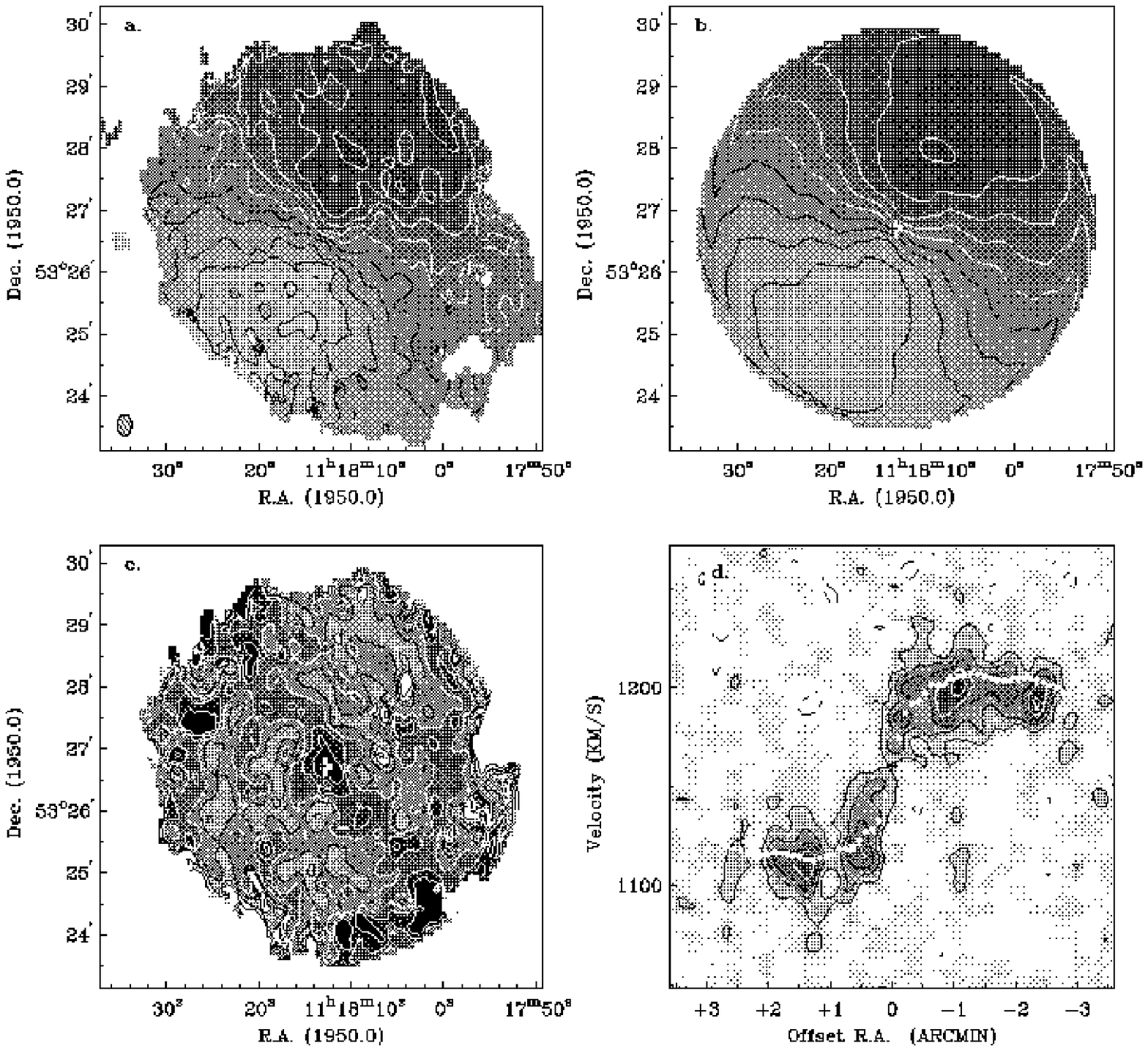}
  \caption{{\it a.} (upper left) \hi\ velocity field of NGC~3631 at
  21\sec$\times$ 14\sec\ resolution. Contour levels are from 1100\,\kms\
  to 1200\,\kms\ in steps of 10\,\kms, where the lower values are found
  to the SE, and the first white contour is at 1160\,\kms. Grayscales
  indicate roughly the same range in velocities. Beam size is
  indicated. {\it b.} (upper right) Model velocity field as determined
  from the rotation curve (see text). Contour and gray levels as in
  Fig.~7a. {\it c.} (lower left) Residual velocity map, obtained by
  subtracting the model (Fig.~7b) from the velocity field
  (Fig.~7a). Contours are at $-10, -6$ and $-2$\,\kms (black) and 2, 6,
  and 10\,\kms\ (white), with grayscales indicating the same range and
  higher values coded darker. {\it d.}  (lower right) Position-velocity
  diagram along the major axis ($\phi=150\deg$) of the 21\sec$\times$
  14\sec\ data set. Contour levels are as in Fig.~1a. Overlaid (white
  dots) is the rotation curve for the whole disc at the same
  resolution. }
\end{figure*}

The velocity field (first moment map) of NGC~3631 at 21\sec$\times$
14\sec\ resolution is shown in Fig.~7a in a contour plus grayscale
representation.  Similar maps were in fact produced at all
resolutions; only one is shown in the present paper.  The velocity
field in Fig.~7a shows a generally regular shape, but a few features
are worth pointing out explicitly.  The shape of the isovelocity
contours indicates that the position angle of the major axis is
practically constant over the disc.  The closed contours representing
more extreme velocity values point out a constant or slightly falling
rotation curve in the outer half of the disc, shown to be indeed the
case in the next paragraph.  At several positions in the disc, notably
about 1\min\ W and $\sim2\min$\ E of the nucleus, and showing up most
clearly along the minor axis, are deviations from the regular shape of
the isovelocity contours that can be recognized as streaming motions
due to a density wave near the \hi\ spiral arms (e.g.  Rots 1975).
From the displacement of the contours one can estimate an amplitude of
$\sim15$\,\kms\ along the line of sight, or $\sim50$\,\kms\ in the
plane of the disc, after deprojection, assuming an inclination angle
of 17\deg\ (RC3, see also next paragraph).  Such values are somewhat
large, but not outside the range of values found in other galaxies for
density wave streaming motions (e.g.  Visser [1980] for M81, Rots et
al.  [1990] for M51, and Knapen et al.  [1993] for M100).

\subsection{Rotation curve}

I derived the rotation curve from the velocity field (first moment maps)
at different resolutions, using a procedure described in detail by
Begeman (1989). In this procedure, the galaxy is divided into a set of
concentric rings, each of which is described by a set of parameters $i$
(inclination angle), $\phi$ (position angle of the major axis), and
$v_c$ (rotational velocity). Additional parameters are the centre of
each ring, and the systemic velocity $v_{sys}$ of the galaxy. These
parameters are fitted using a least squares algorithm. Data points
within each ring are weighted by $|\cos(\theta)|$, where $\theta$ is the
azimuthal angle from the major axis. Data points within 20\deg\ from the
minor axis are excluded from the fits.

NGC~3631 is a galaxy with a low inclination angle ($i=17\deg$, de
Vaucouleurs et al. 1991) which makes it impossible in practice to fit
both $v_c$ and $i$. Rather than assuming a constant rotational
velocity throughout the disc, a constant inclination angle was
used. Using the value of $i=17\deg$ (RC3) results in values for the
rotational velocity of $v_c\sim140$\,\kms, which agree quite well with
the synthetic rotation curve given by Rubin et al. (1985) for a galaxy
like NGC~3631. Rubin et al. give a value of $v_{rot}=132$\,\kms\ for
an Sc galaxy of $M_B=-20$\,mag, the absolute magnitude of NGC~3631
assuming $D=15.4$\,Mpc and the value for $m_B$ from the RC3. The
agreement is surprisingly good given the high uncertainty in the
determination of an inclination angle of only 17\deg. Note that the
use of a different value for $i$ does not change the shape of the
rotation curve, but only scales the values for $v_c$ (larger $v_c$ for
smaller $i$).

Starting out with the 60\sec$\times$ 40\sec\ velocity field, I first
fitted the position of the dynamical centre as $\alpha(1950)=\hmsd
11h18m12.75s (\pm0\fs07), \delta(1950)=\dmsd 53d26m43.0s (\pm$1\sec)
by fixing $i$, $\phi$ and $v_{sys}$ at reasonable values. This
position was determined from the part of the disc at radii larger than
40\sec, where the errors in the fit were acceptable. The dynamical
centre coincides with the optical centre of the galaxy (as given in the
RC3) in declination, but is offset from it by 0.55 seconds of time, or
some 5 arcsec, in right ascension. As a second step, the position of
the centre and the inclination angle were fixed, which resulted in a
satisfactory fit to the systemic velocity of the galaxy,
$v_{sys}=1155.7\pm0.8$\,\kms. This is in good agreement with the
values determined from single-dish \hi\ observations
(e.g. $v_{sys}=1156\pm2$\,\kms, Staveley-Smith \& Davies 1988).

The final rotation curve fit for the whole disc was now made fixing the
position of the centre and $v_{sys}$ at the values described above, $i$
at 17\deg, and fitting $\phi$ and $v_c$ at radial points with 20\sec\
(similar to half beam) spacing. Fits for the receding and approaching
halves of the galaxy were also made.

\begin{figure}
  \epsfxsize=8.5cm \epsfbox{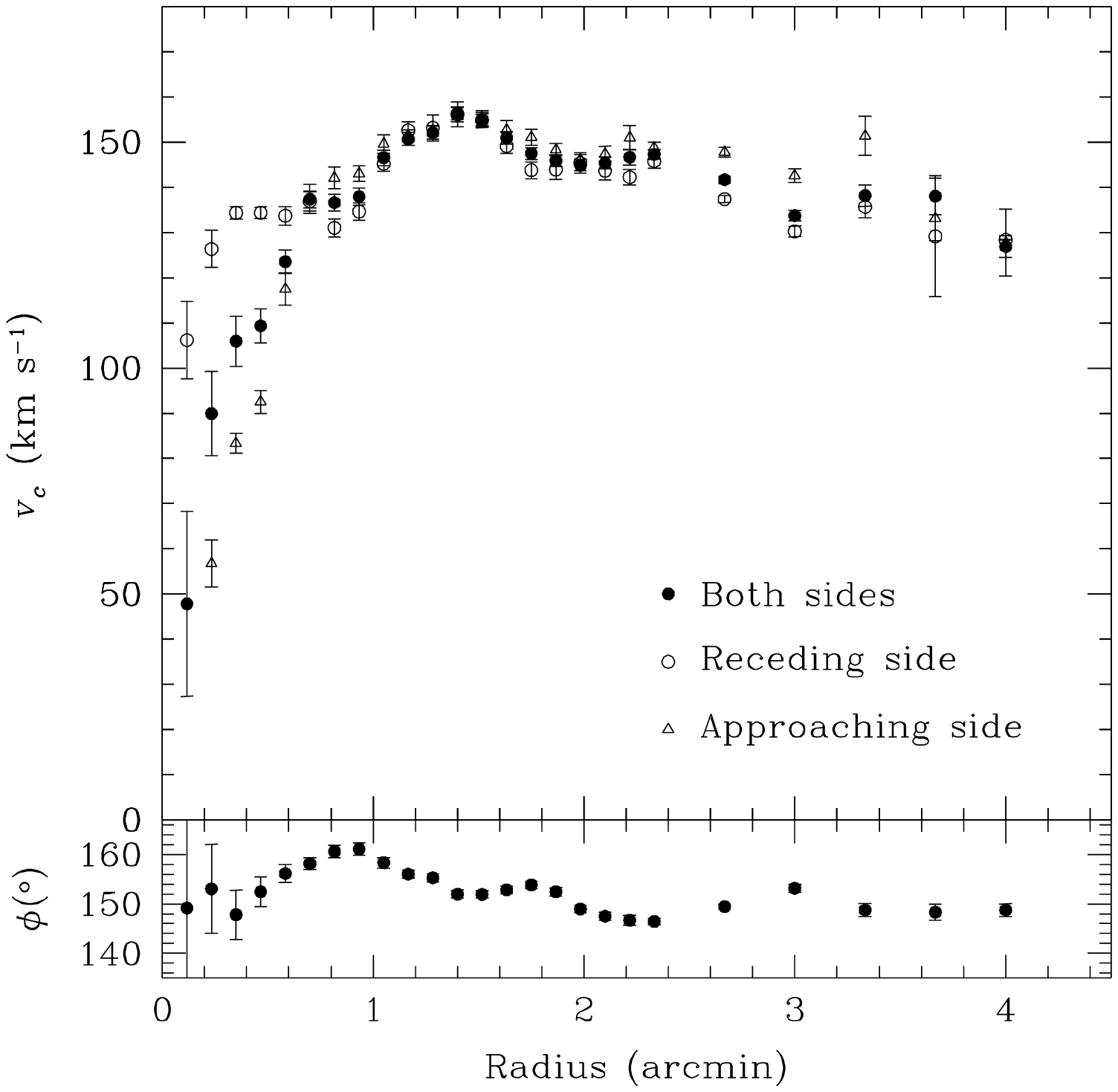}
  \caption{Rotation curve for NGC~3631 (upper panel): whole disc (full
  dots), receding side only (open circles), and approaching side
  (triangles). Formal errors from the least-squares fit are indicated,
  but the uncertainty due to the low inclination of the galaxy is
  larger. Fitted position angle of the major axis (N over E) is shown
  as a function of radius in the lower panel. For both panels, points
  from the centre out to \mind 2.4\ are derived from the
  21\sec$\times$ 14\sec\ resolution velocity field, points at $R>\mind
  2.5$ are from the 60\sec$\times$ 40\sec\ velocity field.}
\end{figure}

Rotation curves for the whole disc, and for receding and approaching
halfs separately, were then fitted for all other resolutions, using as
width of the rings always half the (minor axis) beam size. Following
exactly the same procedure as described for the 60\sec$\times$ 40\sec\
velocity field, values for the dynamical centre and systemic velocity
were first checked, and $\phi$ and $v_c$ subsequently fitted for all
the other cubes at lower resolution. The resulting final rotation
curve for NGC~3631 is shown in Fig.~8. This Figure is a composite of
the fits to the 21\sec$\times$ 14\sec\ resolution velocity field (for
radii out to \mind 2.4) and to the 60\sec$\times$ 40\sec\ velocity
field (larger radii). The top panel shows the rotation curve for the
whole disc, and approaching and receding sides separately, while the
lower panel shows the run of the fitted position angle of the major
axis (measured N over E) against radius. It is clear that $\phi$ is
essentially constant over most of the disc, at a value of
$\phi=150\deg\pm2\deg$. It reaches significantly larger values (up to
$\phi=160\deg$) only around $R=55\sec$, which is probably caused by
the streaming motions prevalent there (see previous section).

The rotation curve rises slowly out to $R\sim\mind1.4$, after which it
declines somewhat. This decline does however depend critically on the
constancy of the assumed inclination (a chance of only 1 or 2\deg\
would suffice to cause the observed drop). Over most of the disc the
rotation curves for approaching and receding sides of the disc behave
very much like the total rotation curve. Apart from the outer points,
the only deviations are found in the inner $\mind0.6\!\,$. These may
well be caused by the adopted central position, which was determined
mostly from the outer regions of the disc, and from data at
60\sec$\times$ 40\sec\ resolution. An overlay of the (21\sec$\times$
14\sec) rotation curve on a position-velocity diagram along the major
axis (Fig.~7d) confirms that the derived rotation curve is very
reasonable, even though the inclination of the galaxy is low.

\subsection{Model velocity field}

The fitted runs of $v_c$ and $\phi$ for the 21\sec$\times$ 14\sec\
data (whole disc fit; see previous section) were used to construct an
axisymmetric model velocity field.  All other parameters ($v_{sys}, i$
and central position) were kept at the values described above.  The
resulting model velocity field is shown in Fig.~7b.  The
counterclockwise deviation of the isovelocity contours near the outer
edge of the model is due to a pair of high values of $\phi$, which may
be artifacts.  The model looks generally smooth, although it is clear
that the streaming motions discussed before show up to a certain
degree.  The model was subtracted from the velocity field to produce a
residual velocity map, which is shown in Fig.~7c.  Values for the
residual velocity of $|v_{res}|>15$\,\kms\ occur exclusively in the
outer 40\sec\ of the map.  Near the centre, where large velocity
gradients occur, residual velocities are positive, around
$v_{res}=10$\,\kms.  In the disc, the residual velocity field shows a
continuous region of negative values (dark in Fig.~7c), aligned more
or less along the direction of the minor axis.  This feature may well
result from the fact that data points within 20\deg\ from the minor
axis were excluded from the fit leading to the rotation curve.  The
residual velocity map also shows the spiral arms, in the form of
positive residual velocities where streaming motions can be identified
in the velocity field (Fig.~7a).  This is most obvious in the form of
the curved feature in Fig.~7c starting N of the nucleus, curving to
the left (E) and then down (S), although other features in the map can
also be identified to lie near or on the spiral arms (compare Fig.~3).
The absence of a symmetric pattern surrounding the centre confirms
that this galaxy does not have a dynamically important
non-axisymmetric component (such as a bar).

\section{Summary}

\begin{table*}
 \centering
 \begin{minipage}{140mm}
  \caption{Parameters and results for NGC 3631}
\begin{tabular}{l l l}
\hline\hline
Parameter & Value & Notes\\
\hline
Morphological type &		SA(s)c & 1\\
Optical centre (1950) & 	\hmsd 11h18m13.3s, \dms 53d26m43s & 1\\
Dynamical  centre (1950) & 	\hmsd 11h18m12.75s ($\pm$0\fs07), \dmsd
53d26m43.0s ($\pm$1\sec) & 2\\
Systemic velocity (heliocentric)& 1155.7$\pm$0.8 \kms & 2\\
Distance &			15.4 Mpc & 3\\
$m_B$ & 			$10.97\pm0.14$ & 1\\
Optical size ($0.5\times D_{25}$) & \mind 2.5 & 1\\
\hi\ radius & 			\mind 4.0$\pm$\mind 0.2 & 2\\
Inclination & 			17\deg & 1, 4\\
Position angle of major axis &	150\deg$\pm2\deg$ & 2\\
\hi\ flux integral&		51.6 Jy\,\kms & 2\\
Total atomic hydrogen mass &	$2.9\times10^9M_\odot$ & 2\\
\hline
\end{tabular}\\
{\bf note 1:} RC3 (de Vaucouleurs et al. 1991)\\
{\bf note 2:} This paper\\
{\bf note 3:} From the systemic velocity, assuming $H_0=75$\kms\,Mpc$^{-1}$\\
{\bf note 4:} Adopted (see Sect. 4)\\

\end{minipage}
\end{table*}

In this paper, I describe new WSRT \hi\ aperture synthesis observations
of the grand-design late-type spiral galaxy NGC~3631. The highest
spatial resolution of the produced data set is of $\secd 15.2\times\secd
11.2$, which allows resolving the spiral arms from the interarm disc,
but data cubes at several lower resolutions were also produced, with the
lowest resolution described here being $60\sec\times40\sec$. These
\hi\ data will be used in subsequent work for detailed studies of SF
processes in and outside the spiral arms, and of the spiral
structure. The main parameters for NGC~3631, as taken from the
literature or determined in the present paper, are listed in Table~3,
and the main results can be summarized as follows:

\begin{enumerate}

\item \hi\ is detected all over the disc of the galaxy, with good
detections, even at the highest resolution, in the interarm  regions and
in the centre. The \hi\ generally follows the spiral arm pattern as seen
in optical images.

\item The \hi\ extends to about $1.5\times R_{opt}$, but most of the
\hi\ is found well within the optical disc. The radial \hi\ profile
peaks just outside $R=1\min$, in the region of the star-forming spiral
arms. The profile shows a slight central depression.

\item Streaming motions due to density waves can be identified near the
locations of the spiral arms in the high-resolution velocity field. They
have maximum amplitudes of $\sim15$\,\kms, or $\sim50$\,\kms when
assuming an inclination angle of 17\deg. No significant other deviations
from axisymmetry are seen in the velocity field.

\item A rotation curve as derived from the velocity field rises slowly
to its maximum at $R\sim\mind1.4$, after which it declines somewhat. The
exact values of the rotational velocity $v_c$ could not be determined
due to the ambiguity between the angle of inclination and $v_c$, but
assuming $i=17\deg$ we find that $v_c\sim140$\,\kms, in general
agreement with synthetic rotation curves for a galaxy like NGC~3631
(Rubin et al. 1985). The position angle of the major axis
$\phi=150\deg\pm2\deg$ is practically constant over the disc.

\item After subtracting a model velocity field, obtained from the
rotation curve, from the original velocity field, the residual map
confirms the existence of streaming motions in the \hi, and the absence
of significant other disturbances in the disc.

\end{enumerate}

\section*{Acknowledgments}

The keen eye of the referee, Prof.  H.  van Woerden, helped improve
the manuscript.  I thank Drs.  K.  Begeman, A.H.  Broeils, R.P.
Olling and D.  Sijbring for help with aspects of the data reduction.
The Westerbork Synthesis Radio Telescope is operated by the
Netherlands Foundation for Research in Astronomy with financial
support from the Netherlands Organization for Scientific Research
(NWO).  The Jacobus Kapteyn Telescope is operated on the island of La
Palma by the Royal Greenwich Observatory in the Spanish Observatorio
del Roque de los Muchachos of the Instituto de Astrof\'{\i}sica de
Canarias.

\bsp

\label{lastpage}

\end{document}